\documentclass[runningheads]{llncs}
\usepackage[T1]{fontenc}
\usepackage{graphicx}
\newcommand\T[1]{\vspace{4pt}\noindent\textbf{#1} }
\usepackage{enumitem}
\usepackage{pifont}
\usepackage[export]{adjustbox}
\usepackage{graphicx}
\usepackage{caption}
\usepackage{flushend}
\usepackage{subcaption}
\usepackage{algorithm,algorithmic}
\usepackage{rotating}
\usepackage{placeins}
\usepackage{tikz,collcell}
\usepackage{balance}
\usepackage{lipsum}
\usepackage{url}

\newcommand*{\cellseta}{\pgfqkeys{/cell}}
\newcommand*{\myCell}{\cellseta}
\newcolumntype{C}[1]{>{\collectcell\myCell}#1<{\endcollectcell}}
\makeatletter
\tikzset{overlay linewidth/.code=\tikz@addmode{\tikzset{overlay}}}
\cellseta{.unknown/.code={\edef\pgfkeys@temp{\noexpand\cellset{box=\pgfkeyscurrentname}}\pgfkeys@temp}}
\makeatother
\cellseta{
  head/.code=\rlap{\scriptsize\rotatebox{90}{#1}}\hspace*{1.2em},
  myCell/.style={
    draw=black,
    overlay linewidth,
    inner sep=+0pt,
    outer sep=+0pt,
    anchor=center,
    fill={#1},
    minimum size=+1.2em},
  box/.code={%
    \tikz[baseline=-0.5ex]
      \node[/cell/myCell={#1}]{};},
  box/.default=none,
  ./.style={box},
  @define/.style args={#1:#2}{#1/.style={box=#2}},
  @define/.list={%
    technical:cyan,
    chi:   green, bribe:  red,
    buy:          orange, todo:       yellow}
  }
\usepackage[most]{tcolorbox}
\tcbuselibrary{skins,breakable}

\usepackage{pifont}
\usepackage{makecell}
\usepackage{lscape}
\usepackage{threeparttable}
\usepackage{amsmath}
\usepackage{amssymb}
\usepackage{enumitem}
\usepackage{afterpage}

\usepackage{hyperref}
\usepackage{cleveref}
\usepackage{xurl} 
\usepackage{tabularray}
\usepackage{lmodern,babel,adjustbox,booktabs,multirow}
\usepackage{kantlipsum}
\usepackage{caption}
\usepackage{subcaption}

 \newtcolorbox{titlebox}[5]{enhanced,center,colframe=black,colback=white,boxrule={#3},arc={#2},auto outer arc,%
 breakable,pad at break*=5pt,vfill before first,before={
 },before={\par\smallskip},after={\par\smallskip},top=12pt,left=4pt,%
 enlarge top by=2pt,
 fontupper=\small,
 title={\rule[-.3\baselineskip]{0pt}{\baselineskip}\normalsize\sffamily\bfseries #1}, varwidth boxed title*=-30pt, 
 attach boxed title to top left={yshift=-10pt,xshift=10pt}, coltitle=black,
 boxed title style={colback=white,boxrule={#5},arc={#4},auto outer arc}
 }

 \newenvironment{casestudybox}[1]
 {\begin{titlebox}{Script \normalfont #1}{0.5pt}{0.5pt}{1pt}{0.75pt}}
 {\end{titlebox}}
\NewTableCommand\SCC[1]{\SetCell{bg=#1}}

\begin{document}
\title{On the Lifecycle of a Lightning Network Payment Channel}
\author{Florian Grötschla\thanks{The authors of this work are listed alphabetically.}, Lioba Heimbach, Severin Richner and Roger Wattenhofer\\
\{fgroetschla,hlioba,richners,wattenhofer\}@ethz.ch}
\institute{ETH Zurich}

\authorrunning{F. Grötschla, L. Heimbach, S. Richner, and R. Wattenhofer} 
%

%
\maketitle              
\begin{abstract}
The \textit{Bitcoin Lightning Network}, launched in 2018, serves as a \textit{layer 2} scaling solution for Bitcoin. The Lightning Network allows users to establish channels between each other and subsequently exchange off-chain payments. Together, these channels form a network that facilitates payments between parties even if they do not have a channel in common. The Lightning Network has gained popularity over the past five years as it offers an attractive alternative to on-chain transactions by substantially reducing transaction costs and processing times. Nevertheless, due to the privacy-centric design of the Lightning Network, little is understood about its inner workings.
In this work, we conduct a measurement study of the Lightning Network to shed light on the lifecycle of channels. By combining Lightning gossip messages with on-chain Bitcoin data, we investigate the lifecycle of a channel from its opening through its lifetime to its closing. In particular, our analysis offers unique insights into the utilization patterns of the Lightning Network. Even more so, through decoding the channel closing transactions, we obtain the first dataset of Lightning Network payments, observe the imbalance of channels during the closing, and investigate whether both parties are involved in the closing, or one closes the channel unilaterally. For instance, we find nearly 60\% of cooperatively closed channels are resurrected, i.e., their outputs were used to fund another channel.
\end{abstract}
\section{Introduction}

The inception of Bitcoin in 2008 marked the creation of the first decentralized cryptocurrency. While the introduction of Bitcoin permanently impacted the way society regards money and finance, cryptocurrencies such as Bitcoin are also known for their extremely small throughput. To tackle this issue, \textit{payment channels} were introduced~\cite{vranken2017sustainability,nadarajah2017inefficiency,poon2015lightning,DW2015channels,green2017bolt,spilman2013channels,Miller2017sprites}. The idea is that instead of settling every transaction on the Bitcoin blockchain directly, Alice and Bob create a payment channel between each other on the blockchain and lock an amount of BTC in the channel, namely, the \textit{channel capacity}. With the payment channel, Alice and Bob can exchange payments directly. Even more, multiple payment channels together form a \textit{payment channel network} that allows users to route their payments across various channels. Thus, users are not required to set up a channel with every individual they wish to exchange payments with but can take advantage of the existing network of channels. To compensate the owners of channels involved in facilitating a transaction, transactions pay a small fee.

The \textit{Lightning Network} is a payment network implementation on top of Bitcoin. Nodes in the Lightning Network \textit{gossip} with each other to exchange information about the nodes and channels in the network. For example, when Alice and Bob create a payment channel between themselves, they might choose to announce the channel in the network such that other nodes in the network know about this channel and can potentially use it to route their transaction. Thanks to these messages, the size and structure of the \textit{public} network, that is, nodes and channels that announce themselves, is generally well understood. There are currently more than 13,000 nodes with 50,000 payment channels that hold over 70M~USD~\cite{1ML}.

Privacy for payments is a key component of the Lightning Network. When Alice sends a payment to Charlie, the Lightning Network is designed so that no other node should be able to know the source and the target of the payment, even if they were involved in routing the transaction. Thus, little is understood of the network's activity and usage as most transactions are not broadcast on the Bitcoin blockchain but rather kept between the two endpoints of a channel. We show that despite these mechanisms, we can extract information on the usage of Lightning channels by analyzing the traces left in gossip messages from the Lightning network and Bitcoin transactions that manage these channels on the blockchain. We do so by matching transaction outputs with the possible transaction blueprints provided by the Lightning protocol and identifying the code paths used to claim funds from these outputs. This can tell us, among other things, whether a channel was closed cooperatively, if one party tried to steal funds by broadcasting an old state to the blockchain, or if the output of a closed channel was used to open a new one.

\T{Contribution.} In this work, we present an empirical study of the lifecycle and usage of Lightning Network payment channels. Through an in-depth analysis of off-chain Lightning gossip messages and on-chain Bitcoin data, we provide the following insights: 
\begin{itemize}[topsep=0pt,itemsep=0pt]
    \item Our longitudinal study of \textit{channel openings} over time quantifies the number of channels opened, the size of channels, and the proportion of publicly announced channels.
    \item Through an analysis of gossip messages, we reason about the \textit{usage of channels during their lifetime} and find indicators to predict the direction of the net flow of routed payments in a channel.
    \item The traces of a channel's closing transaction further allow us to quantify the sizes of any unsettled Lightning Network payments at the time of the closing. We obtained, to the best of our knowledge, the \textit{first dataset of Lightning payment sizes} comprising 21,168 payments. 
    \item Our in-depth study of \textit{channel closings} reveals the channel imbalances at the closing time and the closing type, e.g., whether the channel was closed unilaterally or cooperatively. 
\end{itemize}

\section{Lightning Network}

The Lightning Network is a layer 2 protocol designed to scale Bitcoin: a network of bidirectional payment channels enables off-chain transfer of Bitcoin. Each payment channel established by two nodes in the network represents an edge in the network and allows them to exchange payments by agreeing on updated channel states. In practical terms, each channel has a fixed amount of Bitcoin known as its capacity, which remains constant throughout its operation. However, the ownership distribution of Bitcoin within the channel can change with each transaction. For example, if node $A$ sends node $B$ an amount $x$ of Bitcoin, the balance on A's side of the channel decreases by $x$, while the balance on B's side increases by the same amount.

The underlying mechanism that enables this balance updating process without requiring on-chain transactions is the creation of off-chain commitments. These commitments are essentially signed transactions that reflect the updated balances of the channel but are not broadcast to the Bitcoin blockchain unless the channel is closed. This off-chain nature significantly reduces the load on the Bitcoin blockchain, enabling faster and cheaper transactions. When a payment is made over the Lightning Network, it can be routed through multiple channels to reach its final destination. This is possible due to the interconnected nature of the network, where multiple channels between various nodes form a complex web. Payments can thus be routed across the network, from the sender to the receiver, through intermediary nodes that facilitate the transaction. Each intermediary node deducts a small fee for forwarding the payment, providing an economic incentive to participate in the network.

Importantly, the Lightning Network enables instant and low-cost transactions. The network is further designed to protect the privacy of transactions. Since transactions occur off-chain, they are not recorded on the Bitcoin blockchain, enhancing user privacy. In addition, the origin and destination of transactions routed through the network are difficult to trace for an observer, adding an extra layer of privacy.

\subsection{Channel Lifecycle}
\label{sec:channel_lifecycle}
A payment channel in the Lightning Network is created through a \emph{funding transaction}, maintained/updated by \emph{commitment transactions}, and closed by a \emph{closing transaction}. Generally, only the funding and closing transactions are validated on-chain. Commitment transactions, on the other hand, are held by the nodes involved in the channel and only posted on-chain when a channel is unilaterally closed by one party. The unilateral closing of a channel leads to a timelocked output for that party's funds. We detail more specifics of the transaction types in Appendix~\ref{app:transactions}. 

\T{Funding Transaction.} A funding transaction is a Pay-to-Witness-Script-Hash (P2WSH) transaction using a specified script for the output, which represents the channel~\cite{poon2015lightning}. Thus, on the Bitcoin blockchain, a Lightning Network channel is represented by a single P2WSH output containing the hash of a 2-of-2 multi-signature scheme as the locking script. We also refer to this as the multisig or channel address. The transaction can generally have multiple outputs, with some of them taking the role of ``change''. The script for the funding transaction is defined as follows: 

\begin{casestudybox}{Funding}
\begin{algorithmic}[1]
\STATE \texttt{2 <pubkey1> <pubkey2> 2 OP\_CHECKMULTISIG}
\end{algorithmic}
\end{casestudybox}

The two public keys correspond to the private keys held by the two channel endpoints, and the output can only be spent when both agree. Importantly, transactions of this kind are not unique to Lightning channel openings~\cite{KapposEmpirical2021} but heuristics to identify private channels have been developed (cf. Section~\ref{sec:data}).

\begin{figure}[t]
    \centering
    \vspace{-10pt}
    \includegraphics[width=1.0\textwidth]{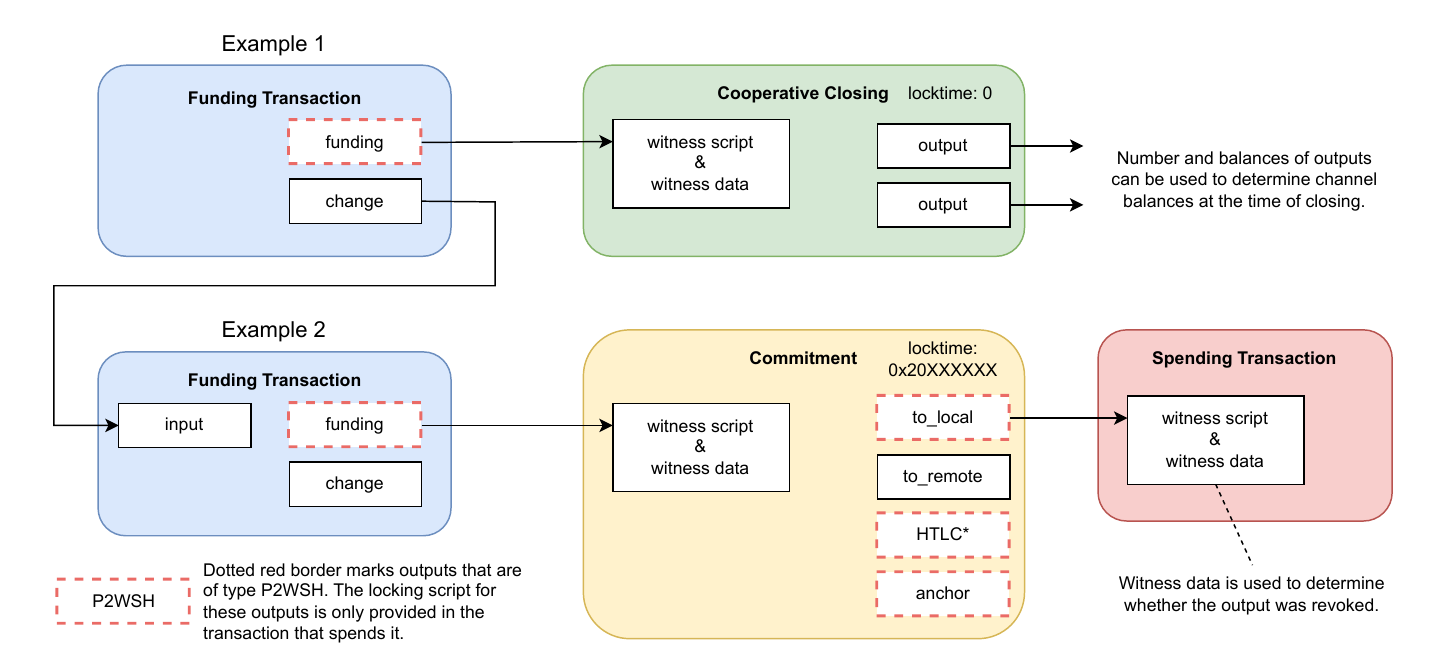}\vspace{-4pt}
    \caption{Two exemplary funding transactions. Cooperative closings spend the 2-of-2-multisig output from the funding transaction and do not have a locktime, while commitments use the locktime field to encode the commitment number. By analyzing the outputs from the commitment, we can classify them into multiple types; some of them are used to send funds to the owner of the commitment (after some timeout), while others represent HTLCs or enable fee bumping. Following the local output for the commitment owner to the spending transaction lets us identify whether the commitment has been revoked. We further analyze whether outputs were used to directly fund other channels. Here, the funding transaction in Example 1 has a change output that funds another channel (i.e., Example 2).}\vspace{-12pt}
    \label{fig:analysis overview}
\end{figure}

\T{Closing Transaction.} A channel is can either be closed \textit{cooperatively} or \textit{non-cooperatively}. If the channel is closed cooperatively, both parties agree on channel balances and jointly decide to close the channel. Both nodes sign a closing transaction that spends the channel funds to their respective wallets. As soon as the transaction is confirmed on the blockchain parties can spend their funds. Otherwise, if the channel is closed non-cooperatively, the party wishing to close the channel submits a commitment transaction to the blockchain. The other node is then given a time window to revoke that transaction (in case an old commitment transaction was submitted that does not reflect the latest status of the channel balances, referred to as \emph{prior state cheating}). If the commitment transaction becomes revoked, all channel funds are awarded to the revoking node as a punishment for not following the protocol. If, however, the time window expires without a revocation, the node can spend the channel funds according to the balances from the submitted commitment transaction. 

\T{Commitment Transaction.} Commitment transactions update the channel balances, and the most recent commitment transaction always represents the current balances between the channel's nodes. These commitment transactions are usually not published on-chain and, thus, allow for fast and inexpensive Bitcoin transfers inside the channel without needing to pay fees on the Bitcoin blockchain. Further, the channel participants sign each commitment transaction. Thereby, invalidating the previous commitment transaction which is essential as it allows for any old commitment transaction to be revoked.

A commitment might be broadcast for various reasons. For example, when one channel party is unresponsive and the other wants to recover its funds. In this case, the broadcaster has to wait for a timeout to pass before they can access their funds -- giving the other party time to invalidate an outdated and replaced commitment. This is referred to as \emph{prior state cheating} and results in all funds being given to the party that invalidated the outdated commitment. If no such invalidation takes place, then after the timeout, the funds can then be accessed by the broadcaster.  

\section{Data Collection and Classification}\label{sec:data}
We collect data from the Lightning Network gossip data as well as the Bitcoin blockchain data. Our data ranges from 1 January 2019 to 23 September 2023, but utilize shortened data ranges for parts of the analysis. In the following, we detail the data collection.

\T{Bitcoin Blockchain Transactions.}
To gather Bitcoin transactions related to channel openings and closings, we utilize the Blockstream Esplora API~\cite{BlockstreamAPI}. In particular, we start with public channels that are announced through gossip messages and retrieve their funding transactions. These are used as a starting point for the private channel discovery and to scan their transaction outputs for usage in later transactions. As most outputs are of type P2WSH as specified in the protocol, the locking scripts are concealed until their usage as a transaction input. Therefore, starting from the funding transactions, we scan all outputs and their usages to detect the closing and spending transactions to infer the type of the transaction output and store further details such as the block height and time the following transaction took place. 

\T{Private Channel Detection.}
While \textit{public} channels announce themselves through gossip messages, \textit{private} channels are never gossiped about publicly. Various heuristics for private Lightning channel detection exist~\cite{NowostawskiEvaluating2019,herrera2019difficulty,tikhomirov2020quantitative}. These are primarily focused on identifying potential funding transactions. We use the following heuristic proposed by Kappos et al.~\cite{KapposEmpirical2021} to identify these channels and calculate associated statistics: 
\begin{enumerate}[topsep=0pt,itemsep=0pt]
    \item We apply the ``Property Heuristic'' to identify Bitcoin transactions that are likely used as Lightning funding transactions. The heuristic includes checking the number, size, and kind of transaction outputs, as well as their compliance with the Lightning specification.
    \item To identify private channels, we employ the ``Tracing Heuristic'' that detects ``peeling chains'' -- sequences of channel opening and closing transactions that are linked within the Bitcoin transaction graph. This heuristic tracks the flow of funds by following the closing and change outputs of channel funding transactions to determine if they are reused in subsequent channel funding transactions. Such reuse suggests that a single entity is involved in both channels. The heuristic can also be applied in reverse to trace the origins of the funding inputs. Channels identified through this method that do not appear in the Lightning Network's gossip protocol data are classified as private, as they are not publicly announced.
\end{enumerate}
Given that we lack information about whether these channels map to any nodes in the publicly accessible network, we limit ourselves to deriving statistics based solely on on-chain data. We discuss the ethical considerations related to private channel detection in Appendix~\ref{app:ethics}. 

\T{Transaction Output Classification.} To deduce the output type, we evaluate the locking script and cross-reference it with known output types within the Lightning specification~\cite{lightningbolts3}. These types encompass \emph{local} outputs, which represent funds time-locked for the commitment owner, \emph{remote} outputs, allocated to the other channel party for direct spending, HTLCs designed for non-confirmed transactions, and anchors enabling fee bumping. In the case of local outputs, we further investigate the path employed for script unlocking, enabling the assessment of potential revocations in instances of prior state cheating. Outputs that remain unspent are categorized as \emph{unspent}, as their output type cannot be inferred without a spending transaction that provides the witness data.

\section{Channel Lifecycle}
Our analysis details the life of a Lightning channel. We commence with the channel opening (cf. Section~\ref{sec:open}), comment on the channel activity during its lifetime (cf. Section~\ref{sec:liftime}), and conclude with the channel closing (cf. Section~\ref{sec:closing}).

\subsection{Channel Opening}\label{sec:open}

\begin{figure}[t]
\centering

\begin{minipage}[b]{.48\linewidth}
    \includegraphics[scale=0.95,right]{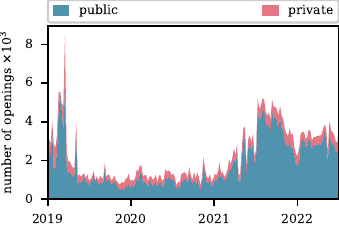}\vspace{8pt}
    
    \caption{Weekly number of public and private channel openings. }
    \label{fig:channelOpening}\vspace{-12pt}
\end{minipage}
\hfill
\begin{minipage}[b]{.48\linewidth}
    \includegraphics[scale=0.95,right]{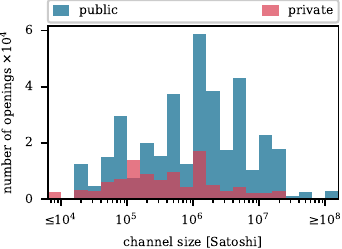}\vspace{-5pt}
    \caption{Private and public channel opening sizes.}
    \label{fig:openingSize}\vspace{-12pt}
\end{minipage}
\end{figure}

The life of a channel begins when its funding transaction is created. In Figure~\ref{fig:channelOpening}, we show the weekly number of public and private channel openings. In 2020, there were consistently around 5,000 channel openings per month. There is a notable increase in 2021, reaching 15,000 monthly openings, ahead a slight decline. The increase and subsequent decline in nodes could be related to the adoption of Bitcoin as a legal tender in El Salvador on 5 June 2021~\cite{alvarez2022cryptocurrencies}. Throughout this period, private channels constituted approximately 22\% of all channel openings.
Figure~\ref{fig:openingSize} further visualizes the channel sizes for private and public channels. We consider the amount of Satoshis (1 Satoshi = $10^{-8}$ Bitcoin) locked in these public and private channels over the entire timeframe. These channels vary widely in size, ranging from four to eight digits of Satoshis, which, as of May 2024, one Satoshi is less than a thousandth USD. Interestingly, private channels tend to have lower average volumes. The reasons for this could be attributed to factors such as specific use cases, privacy considerations within the network, or user preferences when engaging in private channel transactions. 

\subsection{Channel Lifetime}\label{sec:liftime}

\begin{figure}[t!]\vspace{-15pt}
    \centering
    \begin{subfigure}[t]{0.48\columnwidth}
        \includegraphics[scale=0.95,right]{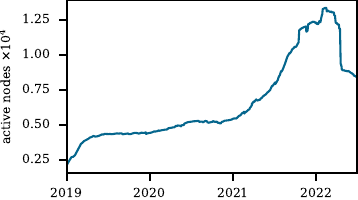}
    \caption{nodes}
    \label{fig:activeNodes}
    \end{subfigure}\hfill
    \begin{subfigure}[t]{0.48\columnwidth}
        \includegraphics[scale=0.95,right]{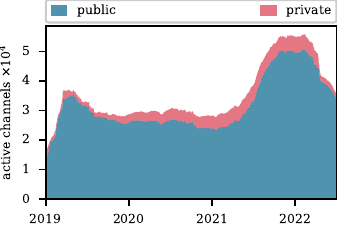}
    \caption{channels}
    \label{fig:activeChannels}
    \end{subfigure}    
    \caption{Number of public nodes (cf. Figure~\ref{fig:activeNodes}), as well as public and private channels (cf. Figure~\ref{fig:activeChannels}) over time.}\label{fig:size}\vspace{-8pt}
\end{figure}

In the following, we focus on the lifetime of the channels. We start by investigating the size of the Lightning Network in terms of the number of active nodes (cf. Figure~\ref{fig:activeNodes}) and the number of active channels (cf. Figure~\ref{fig:activeChannels}). We consider a node to be active if it is involved in at least one open public payment channel. Importantly, for nodes, we only identify public nodes as private nodes are not active in the gossip network. From the start of 2019 until the end of our data period, i.e., 1 July 2022, we observe that the number of active nodes is generally increasing. Notably, there is a significant increase in mid-2021 and a significant drop in the number of nodes in early 2022. Again, we speculate that this could be due to the usage of the Lightning Network in El Salvador. Further, the drop in the number of active nodes in 2022 coincides with an unusually high number of channels closing during that time period (cf. Section~\ref{sec:closing}). We note that the number of active nodes peaked at around 12,500 at the beginning of 2022 and dropped to just over 7,500 by mid-2022.
Similarly, the number of active channels, namely, the number of open channels, is increasing during our data period. However, less so than the number of nodes --- indicating that the average node is involved in fewer public channels in mid-2022 (with four) than at the beginning of 2019 (with six). The number of public channels peaked at over 45,000 in early 2022. For the channels, we also include the number of private channels and observe that the number of private channels is always less than 20\% of the number of public channels. Further, the proportion of private channels peaked in early 2021 and has decreased since then. We further notice a small discrepancy between the proportion of private channel openings (cf. Figure~\ref{fig:channelOpening}) and their proportion of the network. This discrepancy is a result of the short channel lifetime of private channels as we will see in Section~\ref{sec:closing}.

\begin{figure}[t]
\centering
\begin{minipage}[t]{.48\linewidth}
    \includegraphics[scale=0.95,right]{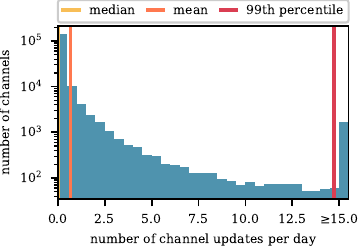}
    \caption{Daily number of channel updates. The median, mean, and 99th percentile are indicated by vertical lines.}
    \label{fig:numberChannelUpdatesPerDay}\vspace{-10pt}
\end{minipage}
\hfill
\begin{minipage}[t]{.48\linewidth}
    \includegraphics[scale=0.95,right]{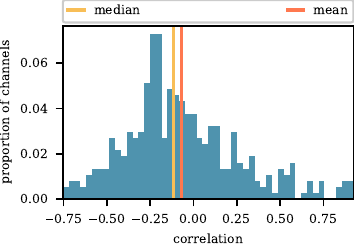}
    \caption{Correlation between proportional fee set by the two channel sides for channels with $\geq$100 updates by each side.}
    \label{fig:correlationProp}\vspace{-10pt}
\end{minipage}\vspace{-5pt}
\end{figure}

\T{Gossip Message Analysis.} We continue by investigating the gossip messages broadcast on the network. Due to gaps in the lngossip~\cite{lngossipDataset} dataset (cf. Section~\ref{app:gossip}), we restrict the following analysis, which depends on these network messages, to a period without gaps, i.e., 1 January 2020 to 1 July 2021. Recall that there are several types of gossip messages, we focus on \texttt{channel\_update} messages here. In more detail, we study the channel updates and analyze whether they give us any insights into traffic patterns in the network. We start with the frequency of channel updates. Every time either channel side adjusts the fees and parameters used for routing, they will broadcast a \texttt{channel\_update} message in the network. Our analysis considers such a message to be an update if the parameters are not identical to the previous message. Figure~\ref{fig:numberChannelUpdatesPerDay} plots a histogram of the mean daily number of channel updates during their lifetime. On average, the channels have 0.69 daily updates, while the median is only 0.05. This discrepancy by a factor of ten between the mean and the median indicates an extremely skewed dataset. That is, there are few channels with many updates and many channels with little to no updates. However, updating channel parameters can be essential to optimize participation in routing. For one, the channel parameters need to be competitive to attract traffic, but just as importantly, the channel parameters are used to avoid the channel becoming depleted by guiding payment flow in the right direction. Note that once a channel active in routing becomes depleted, it is generally closed and reopened, which is costly. Thus, frequent channel updates can indicate that the channel is being actively used in routing transactions through the Lightning Network. We, however, find that only 8.9\% of the public channels update their parameters at least once per day on average and would expect at least one update per day for channels that forward a couple of transactions per day on average. The 99th percentile of channels update their parameters at least 14.7 times a day. We thus believe that these channels actively participate in forwarding transactions through the network. 

\T{Channel Fees.} When updating the fees for routing, nodes specify a \textit{base} (i.e., a flat rate charged per transaction) and a \textit{proportional} (i.e., a rate charged proportional to the transaction size) fee. We will now focus on the proportional fee, as one method of rebalancing a depleted channel is fee management. If one's outbound liquidity is getting low, a strategy is to increase the fees to disincentive nodes from using your outbound liquidity. The opposite could be done at the other channel end. Thus, the proportional fee moving in opposite directions could hint at the net direction of transactions sent through the channel as well as the liquidity imbalance of the channel. To test this hypothesis, we consider all channels with at least 100 updates from either side and plot the correlation between the proportional fee time series from both sides in Figure~\ref{fig:correlationProp}. We find that both the mean and median of the proportional fee correlation across the analyzed channels are negative --- in line with our hypothesis. However, there are also channels with a strong correlation between the proportional fees from either side over time. This could be a sign that both channel sides want traffic regardless of the direction and rely on other rebalancing strategies. 

\T{HTLC Analysis.} Our preceding analysis provides insight into which channels might be involved in routing and the possible direction of flows in channels that we learn by analyzing the gossip messages. However, while the gossip messages allow us to reason about the traffic in the network, they do not offer precise information about the transactions routed through the network. The design of the Lightning Network aims to prevent this information from ever being revealed, but there is an exception during the channel closing. \textit{Hashed timelock contracts (HTLCs)} are the centerpiece of every Lightning Network payment, as they allow for secure and atomic, that is, the entire transaction succeeds or fails, routing through the network. We note that HTLCs are used for both single-hop and multi-hop payments. Importantly, an HTLC represents an unconfirmed transaction, and its size thus corresponds to that of said transaction. In rare cases, these HTLCs are settled on-chain, where the HTLC is not consolidated before the channel is closed. Thus, in these cases, we can observe the size and number of transactions in the channel. 

\begin{figure}[t!]
    \centering
    \begin{subfigure}[t]{0.48\columnwidth}
        \includegraphics[scale=0.95,right]{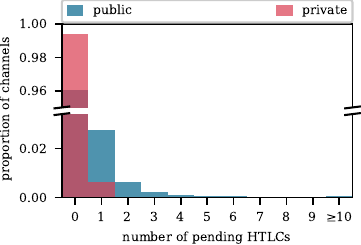}
    \caption{Number of unsettled HTLCs for public and private channels during the closing.}
    \label{fig:pendingHTLCs}
    \end{subfigure}  \hfill
    \begin{subfigure}[t]{0.48\columnwidth}
        \includegraphics[scale=0.95,right]{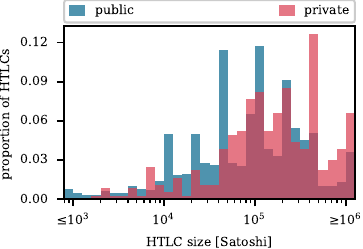}
    \caption{Size of unsettled HTLCs for public and private channels.}
    \label{fig:HTLCSize}
    \end{subfigure}
    \vspace{-2pt}
    
    \caption{Number of unsettled HTLCs (cf. Figure~\ref{fig:pendingHTLCs}) and size thereof (cf. Figure~\ref{fig:HTLCSize}).}\label{fig:htlc}\vspace{-8pt}
\end{figure}

In Figure~\ref{fig:htlc}, we present an analysis of exactly these HTLCs. In total, we observe 20,804 unconfirmed HTLCs in public channel closings and 364 in private channel closings. Figure~\ref{fig:pendingHTLCs} visualizes the number of unconfirmed HTLCs per channel during the closing. For the vast majority of channels, 96\% of public and 99\% of private channels, there are no unconfirmed HTLCs when the channel is closed. For private channels, all remaining channels have precisely one open HTLC. While it is not immediately clear that these are all single-hop payments, it is highly likely to be the case given that none of the 364 private channels with unsettled HTLCs have more than one unsettled HTLC, which is more likely to happen when the channel is involved in routing transactions. 

Finally, these unsettled HTLCs offer a unique insight into the size of Lightning transactions. Figure~\ref{fig:HTLCSize} plots the size of these HTLCs for public and private channels. We start by noting that the HTLCs greatly vary in size and that those unsettled HTLCs we observe for private channels are larger than those in public ones on average. The average HTLC size in private channels is 360,000 Satoshis, whereas it is 230,000 Satoshis in public channels. This could be related to the fact that a larger proportion of unsettled HTLCs in private channels represent single-hop payments. That is, the parties went through the effort of setting up a channel as they were expecting to exchange funds, as opposed to multi-hop payments, where the parties use the existing network to exchange funds. We further notice that there are peaks for the transaction sizes. Many HTLCs are close to round numbers such as 1,000, 2,000, 3,000, or 10,000, indicating that individuals are generally more likely to send payments with these ``round'' sizes through the network. The amounts are usually a few satoshis larger than these multiples, which could be due to fees added on top.

\subsection{Channel Closing}\label{sec:closing}

\begin{figure}[t]
\centering
\begin{minipage}[b]{.48\linewidth}
    \includegraphics[scale=0.95,right]{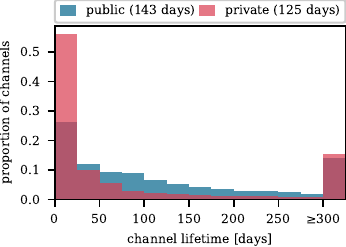}
    \caption{Channel lifetimes of public and private channels. We indicate the mean channel lifetime in the legend.}
    \label{fig:channelLifetime}\vspace{-10pt}
\end{minipage}
\hfill
\begin{minipage}[b]{.48\linewidth}
    \includegraphics[scale=0.95,right]{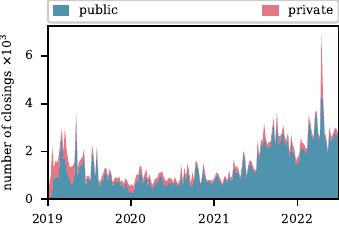}
    \vspace{0.15pt}
    \caption{Weekly public and private channel closings over time. Notice the spike in closings in mid-2022.}
    \label{fig:closings}\vspace{-10pt}
\end{minipage}
\end{figure}

We proceed with an analysis of the end of the channel lifetime: its closing. Figure~\ref{fig:channelLifetime} provides an overview of channel lifetime for public and private instances. Generally, their distribution follows a similar trend. However, extremely short-lived channels make up a more significant proportion of private channels, whereas long-lived channels account for a bigger proportion of public channels. Thus, the average lifetime of public channels is 143 days, which exceeds the average lifetime of private channels at only 125 days. Potentially, some private might have been opened for testing or rebalancing purposes and were thus not announced publicly.

\T{Closing Frequency.} In Figure~\ref{fig:closings}, we plot the weekly number of channel closings for public and private channels over time. While initially, private channels take up a larger proportion of channel closings, the number and distribution of channel closings is relatively stable until mid-2021 with approximately 1,000 channel closings per week. From then on the number of channel closings starts to increase and stabilizes at more than 2,000 weekly channel closings. With one week in mid-2022 exhibiting an abnormally high number of (private) channel closings at more than 6,000. We previously noticed this spike in channel closings due to a drop in the number of nodes and channels in the network at this time (cf. Figure~\ref{fig:size}).

\T{Closing Types.} In the following, we investigate the closing type of channels. Recall that we distinguish between two different types: \emph{cooperative} and \textit{unilateral} (through a commitment transaction). Cooperative closings are bilaterally agreed upon by both channel endpoints through an on-chain transaction. In this case, all funds locked in the channel are directly accessible to both parties. In this case, the number of channel outputs is either one, i.e., all channel funds are with one party, or two otherwise. We will identify the first case as \texttt{coopx1} and the second case as \texttt{coopx2} throughout.

For unilateral channel closings, one party publishes a commitment on the blockchain. The party then has to wait for the passing of a timelock before the funds can be accessed. The timeout allows the other party to publish a revocation if an outdated commitment was published. For closings that were not revoked, we differentiate between three cases by the number of types of outputs: \texttt{local}, \texttt{local + remote}, and \texttt{remote} (cf. Section~\ref{sec:channel_lifecycle}). With \texttt{local}, we identify all unilateral closings, where all funds are with the party that submitted the commitment, and the output has a timelock. With \texttt{remote}, we denote channels that only have one remote output, which does not have a timelock and can be spent immediately by the party that did not submit a commitment. In the case of \texttt{local + remote}, both outputs exist. Finally, we group all revoked unilateral closings as \texttt{revoked} regardless of the number and type of outputs given their sparse occurrence.

\begin{figure}[t!]

    \begin{subfigure}[b]{0.48\columnwidth}
        \includegraphics[scale=0.95,right]{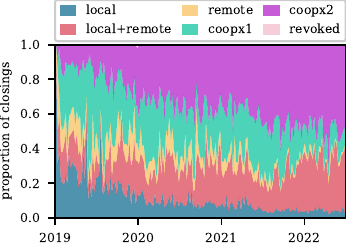}
    \caption{public channels}
    \label{fig:publicClosingsBreakdown}
    \end{subfigure}\hfill
    \begin{subfigure}[b]{0.47\columnwidth}
        \includegraphics[scale=0.95,right]{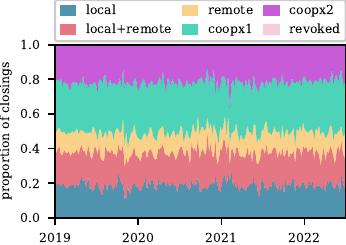}
    \caption{private channels}
    \label{fig:privateClosingsBreakdown}
    \end{subfigure}    
    \caption{Closing outputs for public and private channels. \texttt{local}, \texttt{local + remote}, \texttt{remote}, and \texttt{revoked} are types of unilateral channel closings, \texttt{coopx1}, and \texttt{coopx2} cooperative channel closings. Note \texttt{revoked} closings are extremely rare and thus not visible.}\label{fig:closing}\vspace{-10pt}
\end{figure}

Figure~\ref{fig:closing} visualizes the share of these aforementioned channel closing types for public and private channels, respectively. For public channels (cf. Figure~\ref{fig:publicClosingsBreakdown}), cooperative closings make up the biggest proportion. Together, they account for more than 50\% of all closings, of which cooperative closings with two outputs, denoted as \texttt{coopx2}, are 60\% and those with one output, \texttt{coopx1}, are 40\%. Interestingly, a more significant proportion of channels is closed cooperatively with two outputs than with one at the end of our collection window as opposed to the beginning. Thus, by mid-2022, channels are closed before either side is entirely depleted. Unilateral closings make up slightly less than half of all closings for public channels. For these, the proportion of closing with a single timelocked output, i.e., \texttt{local}, is initially significant and declines over time, whereas those unilateral closings with two outputs, i.e., \texttt{local + remote}, increase over time. With slightly less than 10\% of all closings, unilateral \texttt{remote} closings make up a surprisingly large proportion given that the channel party that will not receive any funds goes through the effort of unilaterally closing the channel. Overall, we notice that by the end of our data analysis period, more public channels are closed before they become entirely unbalanced than in early 2019. Finally, we note that revocations are extremely rare, with a mere 103 observed during our data collection window and thus not visible in Figure~\ref{fig:publicClosingsBreakdown}.

\T{Private Channel Closings.} For private channels (cf. Figure~\ref{fig:privateClosingsBreakdown}), we observe a different pattern. Cooperative closings also make up around 50\% of closings, but the relative increase in those with two outputs cannot be observed. Unilateral closings also account for around 50\% of closings over time. Here, unilateral closings are almost equally split between those with a single \texttt{remote} output and those with a single \texttt{local} output. Furthermore the variations in the relative proportions of channel closings are minimal, especially in comparison to the public channels. Finally, as with public channels, revocations are extremely rare, with 78 occurrences during our time window.  

\T{Closing Balances.} The preceding analysis revealed that public channels are generally closed while no channel end is fully depleted, whereas this is more common for private channels. However, it does not allow us to comment on how unbalanced the channels with two outputs are. In the following, we analyze this by investigating the respective sizes of the channel output(s). We quantify the channel imbalance as follows: 
$ 2\left(\frac{\max\{\text{out}_1,\text{out}_2\}}{\text{out}_1+\text{out}_2}-0.5\right),$
where $\text{out}_1$ and $\text{out}_2$ are the respective channel output sizes. Note that if there is only one output, we set the other output to zero. Thus, a channel is entirely balanced, i.e., both sides have the same balance if our measure is 0, and entirely unbalanced if one side holds all funds when our measure is 1. 

\begin{figure}[t!]
    \centering
    \begin{subfigure}[b]{0.48\columnwidth}
        \includegraphics[scale=0.95,right]{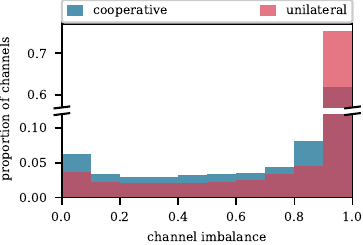}
    \caption{public channels}
    \label{fig:channelBalancePublic}
    \end{subfigure}\hfill
    \begin{subfigure}[b]{0.48\columnwidth}
        \includegraphics[scale=0.95,left]{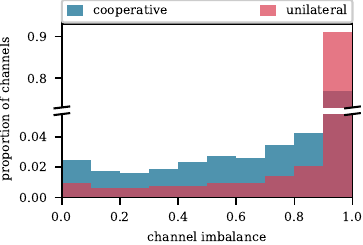}
    \caption{private channels}
    \label{fig:channelBalancePrivate}
    \end{subfigure}    
    \caption{Channel imbalance for public (cf. Figure~\ref{fig:channelBalancePublic}) and private (cf. Figure~\ref{fig:channelBalancePrivate}) channels. A value of 0 indicates that the channel is closed with the funds entirely balanced between the two ends, and 1 indicates that the channel is entirely unbalanced.}\label{fig:balance}\vspace{-10pt}
\end{figure}

In Figure~\ref{fig:balance}, we plot the channel imbalance for public and private channels. We start with public channels, where our previous analysis demonstrated that many channels are closed before one side becomes fully depleted. However, when looking at the channel imbalance in Figure~\ref{fig:channelBalancePublic}, we notice that most channels are extremely unbalanced during their closing. For more than 60\% of cooperatively closed channels and more than 70\% of unilaterally closed channels, one party held at least 95\% of the funds during the closing. On the other hand, only around 5\% of channels are closed in which neither party holds more than 55\% of the channel funds. We further note that unilaterally closed channels are generally more imbalanced during their closing than cooperatively closed ones. One reason for a channel to be unilaterally closed is that one party becomes unresponsive, and this could be the case when they have little to no funds left on their side of the channel and thus become indifferent to what happens with the channel. Thus, this might explain why unilaterally closed channels are more imbalanced than those closed cooperatively. Overall, the average channel imbalance for cooperatively closed channels is 0.79, i.e., one side holds 89.5\% of the channel balance, while the average channel imbalance for unilaterally closed channels is 0.87, where one side holds 93.5\% of the channel funds. 

When we consider private channels (cf. Figure~\ref{fig:channelBalancePrivate}), the trends observed for public channels are even more elevated, channels are even more unbalanced, and even more so for those that are unilaterally closed. More than 75\% of cooperatively closed channels and more than 90\% of unilaterally closed channels have one party holding at least 95\% of the channel funds. Further, less than 2\% of channels, no party holds more than 55\% of the channel funds at the time of the closing. To summarize, the mean channel imbalance for cooperatively closed channels is 0.88, indicating that one side holds 94\% of the channel balance, while the mean channel imbalance for unilaterally closed channels is 0.96, indicating that one party holds 98\% of the channel funds. 

To conclude, regardless of whether the channel is public or private, unilaterally or cooperatively closed, channels are generally very imbalanced once they are closed. Thus, it appears that channels are generally only closed once they can no longer be used to send payments in one direction. A further question is whether the channels are reopened, i.e., whether the channel closing is a means to rebalance the channel on-chain. We find that for 35\% of the closed public channels, at least one of its outputs was used to fund another public channel. In contrast, only 14\% of closed private channels fund another private channel. The reopenings are even more pronounced when only considering cooperatively closed channels. Here, 56\% of closed public channels outputs refund a channel and 33\% for private channels. Additionally, private channels, which are unlikely to be used for routing, are even more imbalanced, indicating that they are potentially created to transfer funds from one side to the other, e.g., to rebalance another potentially public channel or to transfer funds anonymously. 

\section{Related Work}
\T{Lightning Network Topology.} A line of research studies the topology of the Lightning Network. From a theoretical point of view, multiple works study the strategic placement of nodes to route transactions and maximize fee collection efficiently~\cite{davis2022learning,avarikioti2018algorithmic,ersoy2020profit}. Avarikioti et al.~\cite{AvarikiotiRide2020} further game-theoretically study the Nash equilibrium topology of the Lightning Network.  From an empirical point of view,  Seres et al.~\cite{SeresTopological2020} and Lin et al.~\cite{LinLightning2020} present early measurement studies of the Lightning Network topology using Lightning Network gossip messages and comment on high centralization in the network. Subsequent work by Zabka et al.~\cite{ZabkaCentrality2022} takes an in-depth look at the network's centrality to find that the Lightning Network's centrality is increasing. As opposed to investigating the Lightning Network topology, we focus on investigating the lifecycle and usage of the Lightning Network payment channels. Zabka et al.~\cite{ZabkaNode2021} analyze Lightning Network gossip messages to analyze the Lightning Network in further detail. Their work reveals the client implementations used by nodes in the network, as well as their geographical location. Our work combines these gossip messages with on-chain data to investigate the various stages of a channel's lifetime. 

\T{Lightning Network De-Anonymization.} Multiple works have investigated to what extent Lightning Network de-anonymization is possible. Herrera et al.~\cite{herrera2019difficulty} employ probing transactions to unveil channel balances, while Tikhomirov et al.~\cite{tikhomirov2020quantitative} de-anonymize network participants. Romiti et al.~\cite{RomitiCross2021} conduct a cross-layer analysis, combining off- and on-chain data, to de-anonymize participants in Lightning channels. In a similar fashion, Kappos et al.~\cite{KapposEmpirical2021} and Nowostawski et al.~\cite{NowostawskiEvaluating2019} leverage the on-chain data not only to de-anonymize participants but also to identify private channels. We leverage these heuristics to identify private channels and analyze the lifecycles of both public and private channels. Our analysis reveals channel usage patterns, which were previously unexplored.

\T{Rebalancing.} Imbalanced channels are a challenge in the Lightning Network, as they only allow payments to flow in one direction. While the most simple but costly solution to rebalancing a channel is to close and reopen the channel, other (off-chain) rebalancing solutions have been studied and proposed~\cite{papadis2020blockchain,khalil2017revive,conoscenti2019hubs,pickhardt2020imbalance}. Our work reveals insights into the rebalancing methods used by channels. We find that some channels set the routing fees in a manner to attract traffic in the opposite direction, whereas more than half of cooperatively closed channels are reopened, indicating they might have been rebalanced. 

\section{Conclusion}

Previous Lightning Network measurement studies mainly focused on the network topology and network overview statistics, given the privacy protection for transfers in the network. We leverage data leaked through fee updates and on-chain channel closings to extend the understanding of the usage of the Lightning Network by providing further insights into the lifecycle of channels. To the best of our knowledge, our analysis is the first to reveal insights into the usage of private and public payment channels (e.g., routing, rebalancing, etc.) and analyze whether channels are closed cooperatively or unilaterally and possibly reopened. Even more so, we present the first dataset of payments routed through the network -- offering novel insights into the routing activity in the network. We hope that these novel insights into the usage of Lightning channels can guide future developments in the Lightning Network.
%
%
%
\bibliographystyle{splncs04}
\bibliography{references}
\appendix
 
\section{Ethical Considerations}\label{app:ethics}
This work solely utilizes publicly available data: Lightning gossip messages and Bitcoin blockchain transactions. Further, the heuristics used for private channel detection have been previously established in the literature~\cite{NowostawskiEvaluating2019,herrera2019difficulty,tikhomirov2020quantitative,KapposEmpirical2021}. We emphasize that our analysis is conducted on aggregate data, ensuring that no specific private channels or real-world identities can be identified. Furthermore, we deliberately avoid publishing any results or data points that could potentially de-anonymize private channels.

We aim to provide deeper insights into the functioning and performance of the Lightning Network, which is necessary for improving the network while adhering to responsible data usage practices and respecting user privacy.
\section{Bitcoin Lightning Transactions}
\label{app:transactions}

\subsection{Commitment Transactions}
On a technical level, a commitment transaction has a non-zero locktime of the form \texttt{0x20XXXXXX}, where the lower bits are only used to store a concealed commitment number. The locktime does not effectively restrict the time for when the transaction can be mined as it can only take values between 536870912 and 553648127, meaning that the locktime represents a Unix timestamp that can only date back to, at most, 1987. The commitment number itself is obscured by a hash, which makes it computationally infeasible to recover for any outside party. The channel endpoints can, however, verify that the commitment number of a broadcasted commitment transaction corresponds to the most recent one. As the number is obscured, we cannot easily infer the number of commitment updates or balance updates a channel has undergone, even when the commitment transaction was broadcast.

While the commitment transaction spends the P2WSH output of the funding transaction, its output assigns the current channel balances to both parties with their respective outputs. Commitment transactions always have an owner, who corresponds to the channel endpoint that should keep the transaction private until the owner unilaterally closes it. In Lightning nomenclature, the owner of the commitment is referred to as the ``local'' node, while the other channel participant is referred to as the ``remote'' node. Note that remote output can directly be spent, while the local output is timelocked to provide the remote node the opportunity to revoke the output and claim its funds in the case of prior state cheating. The locking script looks as follows~\cite{lightningbolts3}:

\begin{casestudybox}{Local Output}
    \begin{algorithmic}[1]
            \STATE \texttt{OP\_IF} 
    \STATE\hspace{0.5cm} \texttt{\# Penalty transaction} 
    \STATE\hspace{0.5cm} \texttt{<revocationpubkey>} \STATE \texttt{OP\_ELSE}  
    \STATE\hspace{0.5cm} \texttt{`to\_self\_delay`}
    \STATE\hspace{0.5cm} \texttt{OP\_CHECKSEQUENCEVERIFY}  \STATE\hspace{0.5cm}  \texttt{OP\_DROP}  \STATE\hspace{0.5cm}  \texttt{<local\_delayedpubkey>} 
    \STATE\texttt{OP\_ENDIF} 
    \STATE\texttt{OP\_CHECKSIG}
	\end{algorithmic}

\end{casestudybox}

The output can either be spent by the local node once the locktime has expired or directly by the remote node with the \texttt{<revocationpubkey>}. We can classify whether the commitment was revoked or spent by the local node after the timeout by observing which code path was taken in the transaction spending the output.  

In addition to the local and remote transaction outputs, the commitment can have anchor outputs~\cite{lightningbolts3} to allow for fee bumping and HTLC outputs that represent unconfirmed transactions that can be spent after a timeout. Anchor outputs have the following simple form:

\begin{casestudybox}{Anchor Output}
\begin{algorithmic}[1]
\STATE \texttt{<local\_funding\_pubkey/remote\_funding\_pubkey> OP\_CHECKSIG OP\_IFDUP}
\STATE \texttt{OP\_NOTIF}
\STATE\hspace{0.5cm} \texttt{OP\_16 OP\_CHECKSEQUENCEVERIFY}
\STATE\texttt{OP\_ENDIF}
\end{algorithmic}
\end{casestudybox}

HTLCs, on the other hand, are more varied and complex. For the purposes of this paper, we can assume that every output that could not be matched with a local, remote, or anchor output is an HTLC. 

\subsection{Payment Routing}
\begin{figure}[t]
    \centering
    \includegraphics[width=0.9\textwidth]{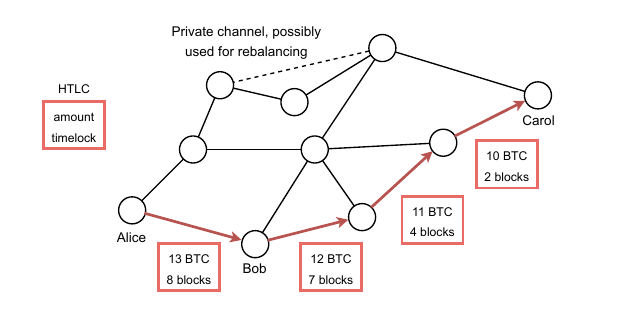}
    \caption{A payment gets routed from Alice to Carol. Carol first creates an invoice to receive 10 Bitcoin. Alice knows the network topology (except for private channels) and chooses a path for the payment (red arrows). Along this path, HTLCs are set up with increasingly larger timeouts towards Alice's end. Nodes on the path request fees to route transactions, in this case, 1 Bitcoin per routed node. Once the HTLCs are negotiated, Carol sends the preimage for the hash backward on the path.}
    \label{fig:routing}
\end{figure}

Setting up a channel between two nodes is expensive and not rational if the two parties only expect to make a single or few payments. However, multi-hop transactions allow Lightning nodes to route their payments over more than just one channel and allow for transferring Bitcoin over a chosen path in the network. This enables nodes to transfer funds to nodes with which they do not have a channel.

To facilitate these transactions, the Lightning Network uses \textit{Hashed Timelock Contracts (HTLCs)}. HTLCs ensure secure transfers by using hashlocks and timelocks. A hashlock has the recipient provide a preimage of a cryptographic hash to claim the funds, while a timelock sets a deadline for the transaction to be completed. An example is depicted in Figure~\ref{fig:routing}.
When the sender, i.e., Alice, initiates a payment, she employs path selection algorithms to determine the most efficient route to the recipient (i.e., Carol). The criteria for path selection typically include minimizing fees, maximizing channel reliability, and ensuring sufficient channel capacity. The sender constructs a potential route by identifying a sequence of intermediary nodes that can forward the payment to the recipient. This involves evaluating the availability and reliability of channels at each hop.

HTLCs are essential to the security and functionality of multi-hop payments in the Lightning Network. These contracts enforce conditional payments based on cryptographic hashlocks and timelocks, ensuring that funds are only transferred if specific conditions are met, thus preventing losses due to misbehaving nodes. The recipient generates an invoice that includes a payment request, a hash of a secret (preimage), and an expiry time. The sender uses this invoice to initiate the payment, creating an HTLC that specifies the hashlock and timelock conditions.
The sender’s node forwards the HTLC to the first intermediary node along the chosen route (i.e., Bob), and each intermediary node, in turn, forwards the HTLC to the next node. This chain of HTLCs ensures that the payment is securely relayed to the recipient. Each HTLC contains the same hashlock, requiring the recipient to reveal the preimage to claim the funds and a timelock that sets a deadline for the transaction. If the preimage is not revealed before the timelock expires, the funds are reverted to the sender.

Upon receiving the HTLC, the recipient reveals the preimage, satisfying the hashlock condition. This preimage is then propagated back along the route: the recipient provides the preimage to the last intermediary node, and each intermediary node verifies the preimage and releases the corresponding HTLC.  Ultimately this process returns the preimage to the sender. Further, the process ensures the atomicity of multi-hop payments; either the entire payment is relayed to the recipient and all nodes in the route receive their respective fees, or the payment fails, and the funds are returned to the sender. This mechanism prevents partial payments and protects against losses due to intermediary node failure or malicious activity.

\subsection{Fees and Incentives}
In the Bitcoin Lightning Network, fees and incentives aim to ensure the network's economic viability and encourage nodes to participate in routing payments. These fees provide the necessary motivation for nodes to allocate resources and maintain reliable channels, thereby enhancing the overall efficiency and stability of the network.

Lightning Network routing fees typically consist of two main components: a base fee and a fee rate. The base fee is a small fixed amount charged by each intermediary node for forwarding a payment, regardless of the payment size. This fee compensates nodes for the basic operational costs for transaction processing. In addition to the base fee, there is a fee rate: a percentage of the payment amount. The fee rate scales with the size of the transaction, providing an additional incentive for nodes to handle larger payments.

The total fee for a multi-hop payment is the sum of the fees charged by each intermediary node along the selected route. For instance, if Alice sends 1000 Satoshis to Dave through intermediary nodes Bob and Carol, each node will deduct their respective fees. Suppose Bob charges a base fee of 1 Satoshi and a fee rate of 1\% (resulting in a total fee of 10 Satoshis), and Carol charges a base fee of 1 Satoshi and a fee rate of 0.5\% (resulting in a total fee of 5 Satoshis). In this scenario, the final amount received by Dave would be 984 Satoshis after deducting the total fees from the initial amount sent by Alice.

Lightning Network nodes set their fees based on various factors, including their operational costs, desired profitability, and competitive positioning within the network. Lower fees attract more routing traffic to a node, increasing its overall revenue through volume. Higher fees, on the other hand, maximize earnings per transaction but could reduce the number of transactions routed through that node. This dynamic creates a competitive environment where nodes balance fee structures to optimize their economic outcomes.

Economic incentives in the Lightning Network extend beyond routing fees. Nodes are also motivated to maintain well-funded and reliable channels to attract users and transactions. High channel reliability reduces the risk of transaction failure, which can undermine user trust and network efficiency. Consequently, nodes invest in robust infrastructure and liquidity management to ensure their channels remain operational and capable of handling a high transactions volume.

Additionally, the Lightning Network incentivizes nodes to participate in the network's growth and stability. As the network expands, nodes benefit from increased routing opportunities and potentially higher revenues. This growth is driven by the network effect, where the utility of the Lightning Network increases with the number of participating nodes and channels, making it more attractive for new users and existing nodes to engage more actively.

\subsection{Gossip Messages}\label{sec:gossip}
In the Bitcoin Lightning Network, nodes communicate through gossip messages that propagate through the entire network. These messages enable nodes to build a local network topology view -- essential for routing multi-hop payments to nodes that are not directly connected. We use three types of gossip messages in our analysis: \texttt{channel\_announcement}, \texttt{node\_announcement} and \texttt{channel\_update}. 

\T{Channel Announcement.}
The \texttt{channel\_announcement} message is broadcast by a node to inform the network about a new channel that it has established. This type of message is crucial for the dissemination of information regarding public channels. Public channels are those that are announced to the network, making them visible to all nodes. In contrast, unannounced channels are considered private channels and are not propagated via gossip messages, thus remaining hidden from the broader network. The \texttt{channel\_announcement} message includes details such as the channel's unique identifier, the public keys of the nodes at both ends of the channel, and proof of the channel's existence on the Bitcoin blockchain. By broadcasting this information, nodes help other participants update their local views of the network topology, facilitating the routing of payments through newly established channels.

\T{Channel Update.}
The \texttt{channel\_update} message informs the network about the parameters of an existing channel. These parameters include the routing fees, time constraints, and any other conditions that might affect the use of the channel for forwarding payments. The Lightning Network is a directed network, meaning that each channel has parameters that must be specified and broadcast by both participating nodes individually. Consequently, each node sends its own \texttt{channel\_update} message to communicate its view of the channel's parameters. This ensures that all nodes in the network can accurately calculate the cost and feasibility of routing payments through any given channel. Regular updates are necessary to reflect changes in fees or channel status, thereby maintaining an up-to-date and reliable network topology.

\T{Node Announcement.}
With the \texttt{node\_announcement} message nodes share additional information about themselves. This message can include metadata such as the node's public key, network addresses, supported features, and alias (i.e., a human-readable name). By broadcasting a \texttt{node\_announcement}, a node enhances its visibility within the network, making it easier for other nodes to identify potential routing partners. This announcement helps nodes build a comprehensive view of the network's participants, facilitating better decision-making when establishing payment routes. Additionally, \texttt{node\_announcement} messages contribute to the network's overall resilience by ensuring that nodes have access to a wide range of information about their peers.

\section{Lightning Network Gossip Messages.}
We obtain historic Lightning gossip messages from Lightning Network Research Topology Dataset~\cite{lngossipDataset}. It contains gossip messages collected and synchronized across multiple nodes. In particular, channel announcements, channel updates, and node announcements are logged. 

\label{app:gossip}

\begin{figure}[t]\vspace{-0pt}
    \centering
    \begin{subfigure}[b]{0.48\columnwidth}
        \includegraphics[scale=0.95,right]{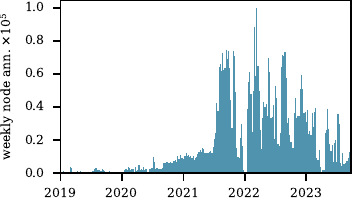}
    \caption{node announcements}
    \label{fig:nodeann}
    \end{subfigure}\hfill
    \begin{subfigure}[b]{0.48\columnwidth}
        \includegraphics[scale=0.95,right]{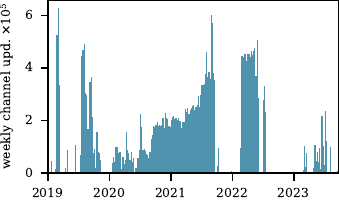}
    \caption{channel updates}
    \label{fig:channelup}
    \end{subfigure}    
    \caption{Daily number of node announcements (cf. Figure~\ref{fig:nodeann}) and channel updates (cf. Figure~\ref{fig:channelup}) recorded in the lngossip~\cite{lngossipDataset} dataset.}\label{fig:gossip}
\end{figure}

We analyze the number of gossip messages recorded in the lngossip~\cite{lngossipDataset} dataset. There are three types of gossip messages, we focus on \texttt{node\_announcement} and \texttt{channel\_update} messages. In Figure~\ref{fig:gossip}, we plot the weekly number of such messages recorded in the lngossip~\cite{lngossipDataset} dataset. We notice a generally increasing trend in the number of messages. Further, there are more channel updates recorded in the network than node announcements. However, this is unsurprising as channel updates are propagated by both ends of the channel anytime they adjust their parameters. Notice that there are some gaps in the dataset, e.g., the end of 2019 or the second half of 2022.

These data gaps are likely attributable to the inherent challenges in data collection rather than any issues within the Lightning Network. Despite these gaps, the lngossip dataset remains the most comprehensive and detailed source of information available on Lightning Network activity. The missing periods can be due to temporary disruptions in the data collection infrastructure, variations in the availability of data collection nodes, or network topology changes that briefly impacted data logging. Nonetheless, the overall trend and volume of gossip messages provide a robust basis for our analysis until July 2022. After this timeframe, the number of channel updates becomes too scarce and does not facilitate a proper analysis. Thus, for the subsequent analysis, we restrict the data period from 1 January 2019 to 1 July 2022. 

\section{Locktimes and Timeouts.}%
\begin{figure}[t]
\centering
\begin{minipage}[t]{.48\linewidth}
    \includegraphics[scale=0.95,left]{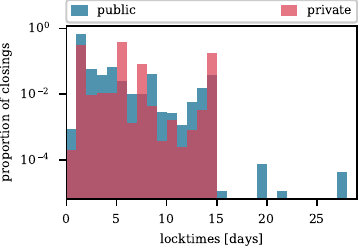}
    \caption{Used locktimes for local outputs in the number of days. The vast majority of outputs have a locktime of 24h (take note of the log scale) or up to 2 weeks.}
    \label{fig:timelock}
\end{minipage}
\hfill
\begin{minipage}[t]{.48\linewidth}
    \includegraphics[scale=0.95,left]{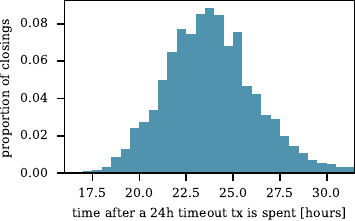}
    \caption{Time after which a local output with a locktime of 144 blocks (exactly 24h at a 10-minute block interval) is spent. }
    \label{fig:time}
\end{minipage}
\vspace{-10pt}
\end{figure}

\label{app:locktimes}%
We further analyze the locktimes of local outputs in commitment transactions. This is the amount of time that the output of the commitment owner is locked for and can be revoked by the other channel party. As is visible in Figure~\ref{fig:timelock}, the majority (about 56\%) of public channels choose a locktime of 24 hours (144 blocks). Apart from this, timeouts up to 2 weeks are also common. Private channels follow this trend, although with the difference that locktimes of 24h and 7 days are equally common. This may hint at the fact that private channel parties put less trust in each other and thus prefer a longer timeframe to revoke an old commitment that tries to steal funds from them.

As locktimes are specified in the number of blocks between commitment transaction and spending of the output, it is not guaranteed that a timeout of 144 blocks can only be spent after 24 hours, as the blockchain might move faster or slower than the average of one block every 10 minutes. We observe in Figure~\ref{fig:time} that this leads to some outputs being spent after considerably less time. This fact should be kept in mind when setting the timeout for a channel, as one might have to react faster to revoke a commitment in the case of prior state cheating or wait longer until the output can be spent. 

\begin{figure}[t]
\centering
\begin{minipage}[t]{.48\linewidth}
    \includegraphics[scale=0.95,left]{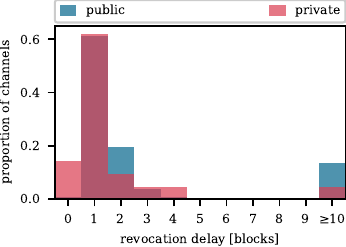}
    \caption{Delay between the publishing of the outdated commitment transaction and the publishing of the revocation on-chain. Some revocations even occur in the same block. }
    \label{fig:revocationDelay}
\end{minipage}
\hfill
\begin{minipage}[t]{.48\linewidth}
    \includegraphics[scale=0.95,left]{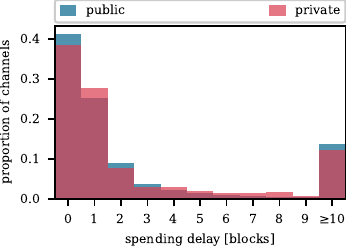}
    \caption{Delay between the expiry of a timelock on the channel output belonging to the party that submitted the commitment transaction unilaterally and the spending of the output.}
    \label{fig:spendingDelay}
\end{minipage}
\end{figure}

\section{Spending of Local Outputs.} We investigate one final aspect of the channel closing --- for unilateral closings, how long does it take for revocations to be published if they occur, and how long does it take until timelocked channel outputs are spent once their lock expires? In Figure~\ref{fig:revocationDelay}, we plot the revocation delay as well as the number of blocks between the publishing of the commitment transaction and the publishing of the revocation, for the 181 revocations in our dataset. Notice that for both public and private channels, the revocation delay tends to be very small. In particular, 62\% of revocations for public channels are posted within one block, whereas 76\% of revocations for private channels are within one block. Remarkably, 14\% of private channel revocations occur in the same block, so someone reacts to a transaction that is waiting for inclusion in the \textit{mempool}, the public waiting area for transactions. We reiterate that revocations are extremely rare, indicating that, in general, channel participants behave well and do not publish outdated commitments. Further, when revocations occur, they are almost immediately posted on-chain in the majority of cases.

Figure~\ref{fig:spendingDelay} further plots the spending delay, i.e., the number of blocks between the timelock expiration on outputs belonging to the party that closes the channel unilaterally and the subsequent spending of the output. We start by noting that the distribution of the spending delay is very similar for private and public channels. In around 40\% of cases, for public and private channels, the previously locked output is spent in the same block as the lock expires, and in around 25\%, one block afterward. Potentially, given how quickly the previously locked outputs are spent, transactions wishing to spend the output are already waiting in the mempool ahead of time and included by a miner once the lock has expired. For both public and private channels, less than 10\% of the time, it takes at least 10 blocks for the funds to be spent.

\end{document}